\documentclass[a4paper,twoside]{article}

\usepackage{epsfig}
\usepackage{subcaption}
\usepackage{calc}
\usepackage{amssymb}
\usepackage{amstext}
\usepackage{amsmath}
\usepackage{amsthm}
\usepackage{multicol}
\usepackage{pslatex}
\usepackage{apalike}
\usepackage{hyperref}
\usepackage{soul}
\usepackage{color}
\usepackage{tabularx}

  \usepackage{wasysym}
  \usepackage{booktabs}
  \usepackage{array}
  \usepackage{threeparttable}
  \usepackage{multicol}
  \usepackage{blindtext}
  \usepackage{graphicx}
  \usepackage{caption}
  \usepackage{multirow}
\usepackage[bottom]{footmisc}
\usepackage{SCITEPRESS}     

\begin{document}

\title{Design, Implementation, and Evaluation of Blockchain-Based Trusted Achievement Record System for Students in Higher Education}

\author{\authorname{Bakri Awaji\sup{1}\sup{,}\sup{2}, Ellis Solaiman\sup{1}}
\affiliation{\sup{1}School of Computing, Newcastle University, Newcastle upon Tyne, UK}
\affiliation{\sup{2}Collage of Computer Science and Information Systems, Najran  University, Najran, Saudi Arabia}
\email{b.h.m.awaji2@newcastle.ac.uk, ellis.solaiman@newcastle.ac.uk}
}


\abstract{With a growing number of institutions involved in the global education market, it has become increasingly challenging to verify the authenticity of academic achievements such as CVs and diplomas. Blockchain is an enabling technology that can play a key role in solving this problem. This study introduces a blockchain-based achievement record system that produces a verifiable record of achievements. The proposed system aims to facilitate the process of authentication and validation of certificates reliably, easily and quickly, leveraging the unique capabilities offered through Blockchain technology (public Ethereum Blockchain) and smart contracts. We present the design and implementation of the system and its components and tools. We then evaluate the system through a number of studies to measure the system’s; usability, effectiveness, performance, and cost. A System Usability Scale (SUS) test gave a scale of 77.1. Through a literature survey we demonstrate that this system is a significant improvement on legacy systems, being both more user-friendly and more efficient. We also conduct a detailed cost analysis and discuss the positives and limitations of alternative blockchain solutions. }

\onecolumn \maketitle \normalsize \setcounter{footnote}{0} \vfill

\section{\uppercase{Introduction}}
\label{sec:introduction}
Every higher education student, for example any individuals in tertiary education studying to complete an academic degree \cite{yumna2019use}, must have a university learning record in which their university progress is documented. The higher education system adopts a largely unified approach to creating these records and providing official transcripts to validate the students’ academic achievements. A student is given proof of their performance through an official university transcript \cite{yumna2019use}. Official transcripts are important records for determining an individual’s employment because they allow potential employers to check their candidate’s education. Moreover, employers may evaluate a candidate’s skills and suitability for the job by asking them to submit a work portfolio. Research indicates that work opportunities are significantly enhanced through the provision of adequate achievement record (service-based or project-based)~\cite{han2018novel}\cite{watters_blockchain_2016}. Nonetheless, without a reliable achievement-recording framework, it is difficult to guarantee that transcripts, or work provided in a portfolio, are genuine works by the candidate. Thus, reliable learning records can be incredibly valuable. At present, most people use traditional methods when applying for jobs, such as the provision of a resume or CV. There are a number of online sources that can be used to help individuals to create CVs, and various structures and styles can be employed. Social networking sites including Facebook and LinkedIn can also be valuable platforms when it comes to creating CVs.

There are no methods established to date that can allow employers to validate the achievements claimed on a candidate’s CV~\cite{cappelli2019your}. Research performed by the Higher Education Degree Datacheck  (2021) revealed that around 30\% of students and graduates fabricated or exaggerated their academic achievements~\cite{higher_education_degree_datacheck_higher_2021}. Thus, this is a great concern faced by employers. NGA HR services are a well-known human resources business process outsourcing (BPO) company in the UK. They revealed statistics to show that 90\% of HR directors had witnessed exaggerations on a job application~\cite{nga_human_resources_nga_2021}.
Moreover, there are several factors associated with achievement records and CVs that can cause a lack of trust, \cite{jirgensons2018blockchain} one of which is poor data continuity. For the most part, learning data remains static, even if students transfer to a different institution. Each institute has its own independent Learning Record Stores (LRSs), meaning that data gathered at previous facilities are unable to be analysed. This generates a cold-start issue, in which there is insufficient data held by the current institution to efficiently customize and monitor the progress of their students~\cite{jirgensons2018blockchain}.

Employers and various other authorities have significant concerns regarding the validation of academic certificates for a number of reasons. For example, some institutions are no longer operational or fail to maintain accurate records. These cases pose significant challenges when it comes to validating the authenticity of educational certificates. More and more institutions are becoming involved in the global education market, and this furthers the difficulty of keeping up-to-date with certificate verification~\cite{vidal2019analysis}. Moreover, a study performed by Han~\cite{henle2019assessing} and Vidal~\cite{vidal2019analysis} showed that, on average, companies spend as much as £40,000 per year to address these issues. Fraudulent achievement records cause major issues for employers and other, honest candidates who are unable to compete with dishonest candidates. It is thus crucial to develop and implement effective measures to prevent certification fraud. 

Research performed by the NGA (2018) highlighted a number of common areas in which candidates fabricated or lied about the information on the achievement records. The findings showed that 44\% of candidates exaggerated or fabricated their achievement information, while 43\% lied about their work history. Moreover, 39\% of participants lied about their professional qualifications and 32\% about their education qualifications. Additionally, 27\% falsified their membership to an industry body, whilst 24\% provided false references~\cite{nga_human_resources_nga_2021}\cite{sanmogan_how_2018}. Poor recording and validation standards of students’ non-academic achievements is also a further matter of concern because these cannot be verified on official transcripts. Thus, it is impossible to verify such activities, including extra-curricular activities, prizes and employability awards, as well as voluntary work and positions in student union clubs and societies. Most studies that have investigated CV fraud, have found the topic to have a significant negative effect~\cite{henle2019assessing}.

Blockchain technology may play a significant role in addressing the issues outlined above. Blockchain technology has a number of immutability and security features that have encouraged researchers to explore its possible use in various domains, including cloud computing, banking, IoT and education. One key advantage of blockchain technology is that smart contracts can be programmed to automate data storage and validation processes~\cite{han2018novel}. 
A smart contract can be defined as an event-condition-action stateful computer program that can be used on top of blockchain to create a distributed application that can be used by numerous parties who are unable to trust each other~\cite{yumna2019use}.  In Molina-Jimenez et al., the key concepts of smart contracts and their use are discussed in greater detail~\cite{molina2018implementation}.

The implementation of a blockchain-based achievement record system would be highly beneficial for students, employers and higher education institutions as it could enable a verifiable record of achievements to be documented. Through this system, students would be able to showcase their achievements to potential employers, which in turn improves their employability. Moreover, official transcripts enable students to assess their progress and plan for their future careers (both individually and with external party support). This ultimately helps them to enhance their extra-curricular and non-academic skills. Moreover, knowing that their academic achievements are documented in a transcript can encourage students to work hard and maintain strong work records, which adds further value to their higher education experiences. A trusted and reliable achievement recording system would also benefit the education system itself because it can reduce administrative tasks. It may also improve the quality standards of student admissions by making students’ achievements transparent. Moreover, such systems would also have advantages for employers, including the provision of reliable and verified achievement records. It would also enable employers to gain a full, detailed picture of a candidate’s higher education achievements, which would be advantageous for the recruitment process.

This paper aims to introduce a blockchain-based achievement record system that generates a verifiable record of achievements for students in higher education. The proposed system aims to simplify and expedite the certificate authentication and validation process by exploiting the unique capabilities provided by Blockchain technology (public Ethereum Blockchain) and smart contracts. This paper describes the system's design and implementation and its components and tools. We then evaluate the system's usability, effectiveness, performance, and cost through a number of studies. 

The remainder of the paper is organised as follows: Section \ref{RELATED WORK} presents the related work, while section \ref{SYSTEM DESIGN AND IMPLEMENTATION} discusses the design and implementation of the proposed system. Then, section \ref{EVALUATION} presents the evaluation, and finally, in section \ref{CONCLUSIONS}, the conclusion.

\section{RELATED WORK}
\label{RELATED WORK}
Several solutions have been suggested and implemented in response to the highlighted issues; as table \ref{Features} shows. 
OpenBadge \footnote{\url{https://openbadges.org}}  is a standard for digital credentials established by Mozilla and is now controlled by the IMS Global Learning Consortium.  In open badges, a wallet is created for participants to add the certificates as badge. The entity issuing the certificate will have provided authentication of it automatically via the wallet, before it being available. The certificate’s inclusion in the retained record of the wallet only occurs once the legal authentication procedure has been successfully completed. Therefore, the wallet’s certificate list offers veracity for all entities and individuals.  Fundamentally, issues such as damaged or misplaced physical certificates and their associated management and printing expenditure both financially and time-wise have been tackled through OpenBadges. Nevertheless, prospective single-point failures in the service or database of OpenBadges pose safety, security, dependability and transparency issues relating to the platform’s management of the issued certificate database. As a result, entities particularly state institutions that feel they lack control over the database provided by OpenBadge have shown reluctance and scepticism over the platform’s adoption~\cite{virkus2019use}.

An alternative program for the issuing and validation of certificates that relies on Blockchain is the cutting-edge, Massachusetts Institute of Technology (MIT) project called Blockcerts \footnote{\url{https://www.blockcerts.org}} \cite{schmidt2016blockcerts}. Blockchain can be used to develop programs for issuing and validating certificates through Blockcerts, which comprises decentralised and open mobile applications, tools and databases. Various documents, including practice permits, education certificates, or criminal records, can all be incorporated. Despite the issuer of the certificate and any other entity not being involved in the validation process, the certificates’ dependability is still guaranteed by Blockcerts. Moreover, if Bitcoin continues to exist, then single-point failures will not be a problem, according to Blockcerts. As a result, continuous accessibility and executability of 
Blockerts’ services have been achieved. The activities of all users are entirely their own responsibility because complete user privacy is promoted by Blockcerts. One instance is how certificate issuance is permitted solely by issuing entities. The issuing entity and the owner of the certificate both have to consent to a certificate’s rescindment, with the irrefutability of such activities being ensured. The certificate is represented by the transactions managed by every Bitcoin address, meaning that a straightforward process is offered by Blockcerts. Accordingly, customers may be persuaded to adopt the program, while its openness is also enhanced. Additionally, mobile software may be used to undertake every certificate management function. Nevertheless, the implementation of Blockcerts is confronted with a number of difficulties \cite{nguyen2018cvss}. CVTrust \footnote{\url{https://www.cvtrust.com}} , Smart Diploma \footnote{\url{https://smartdiploma.io}} , as well as Block. co \footnote{\url{https://block.co}}  are further applications that engage in certificate issuance and validation on the basis of Blockchain. Nevertheless, their adoption approach and technical resolutions are not thoroughly clarified.
\begin{table}[ht] \small
    \centering
	\caption{Summary Comparison of Various Solutions.}
    \label{Features}
  \begin{tabular}{|p{2cm}|c|c|c|c|c|c|c|}
    \hline 
    Systems & \multicolumn{7}{c|}{Features}\\
    \cline{2-8}
   
     &\rotatebox{90}{Accreditation}&
     \rotatebox{90}{Verification}&
     \rotatebox{90}{Privacy}&
     \rotatebox{90}{Transparency}&
     \rotatebox{90}{User Experience}&
     \rotatebox{90}{Accessibility}&
     \rotatebox{90}{Sharing Record}
     \\
\hline
 OpenBadge& $\bullet$  & $\otimes$ & $\bullet$  & $\oslash$  & $\bullet$  & $\bullet$  & $\otimes$ \\ \hline
 Blockcerts    & $\otimes$  & $\bullet$ & $\bullet$  & $\bullet$  & $\bullet$  & $\bullet$  & $\bullet$ \\ \hline
Block.Co     & $\bullet$  & $\bullet$ & $\bullet$  & $\otimes$  & $\bullet$  & $\bullet$  & $\otimes$ \\\hline
Smart Diploma& $\otimes$  & $\bullet$ & $\bullet$  & $\otimes$  & $\bullet$  & $\bullet$  & $\otimes$ \\\hline
 CVTRUST       & $\otimes$  & $\otimes$ & $\bullet$  & $\oslash$  & $\bullet$  & $\bullet$  & $\oslash$ \\\hline
 CVSS          & $\bullet$  & $\bullet$ & $\bullet$  & $\bullet$  & $\bullet$  & $\bullet$  & $\otimes$ \\\hline
 Our System    & $\bullet$  & $\bullet$ & $\bullet$  & $\bullet$  & $\bullet$  & $\bullet$  & $\bullet$ \\\hline

\end{tabular}
$\oslash$  partially provided, $\bullet$  provided, $\otimes$  unprovided
\end{table}
 
\section{SYSTEM DESIGN AND IMPLEMENTATION}
\label{SYSTEM DESIGN AND IMPLEMENTATION}

\begin{figure*}[!htbp]
    \centering
    \includegraphics[width=0.8\textwidth]{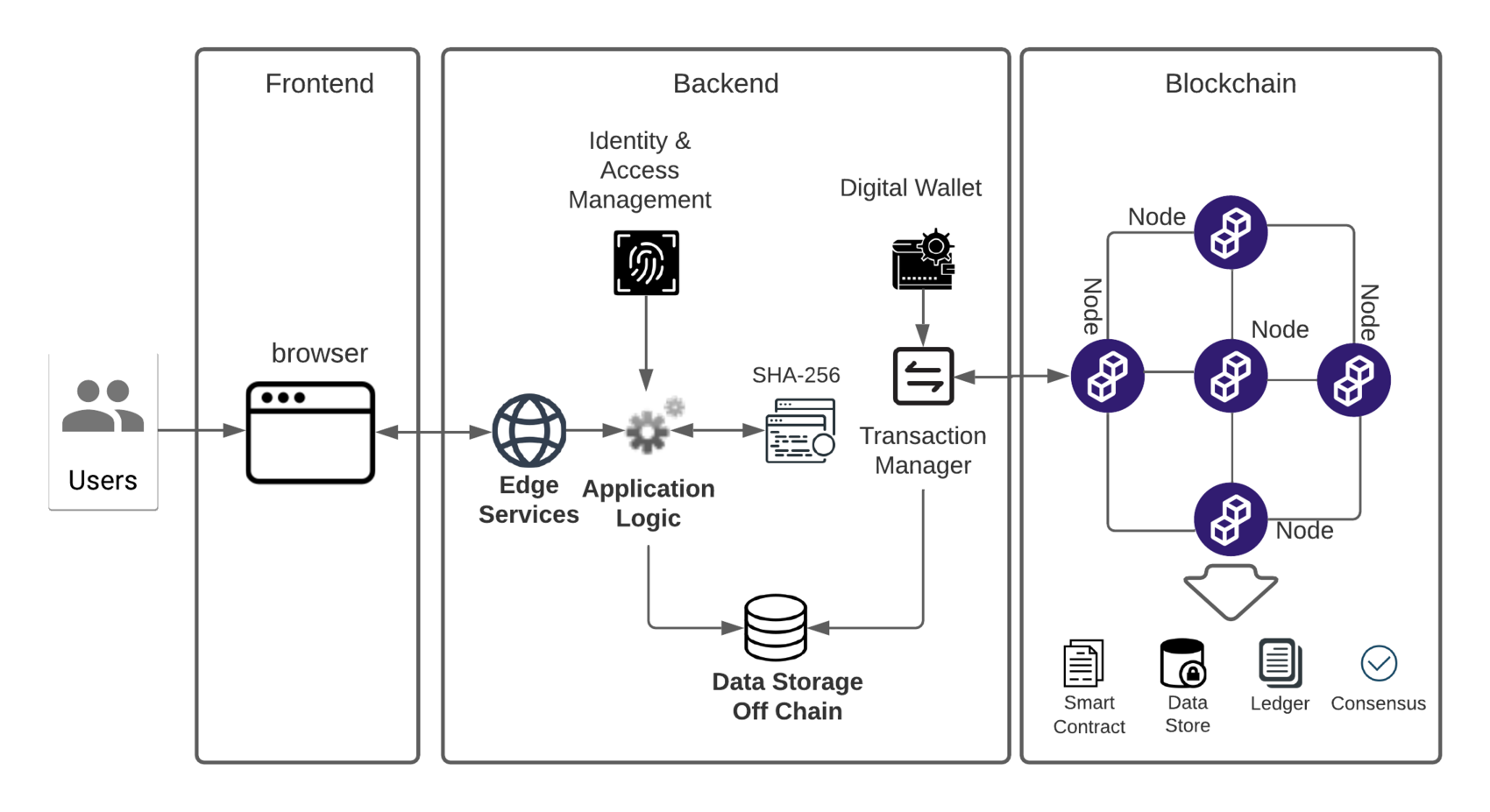}
    \caption{The System Structure.}
    \label{fig:Figure 1:The System Structure.}
\end{figure*}

A trusted achievement record is a secure system that aims to record and authenticate certificates, key learning activities, and achievements. The system’s conceptual model is designed by gathering important information on stakeholders’ thoughts and outlooks on an achievement record system that uses blockchain and smart contract as described in Awaji et al.\cite{awaji2020investigating} and \cite{awaji2020blockchainS}. 

Figure \ref{fig:Figure 1:The System Structure.} illustrates the overall system design and demonstrates the requirements and components, including a frontend and backend as described in Awaji et al.\cite{awaji2020blockchainD}. This architectural design means two ends (frontend and backend) with distinct set dependencies, known as libraries and frameworks. While the frontend acts as a presentation layer that the end-user is introduced with upon entering the site, the backend provides the data and logic which enables the frontend to function.
The frontend is designed to display web pages on PC, Tablets, or Smartphones and contains components that perform different functions in the system. For example, APPLICATION LOGIC component which controls the interfaces of the system and their contents based on the type of user, IDENTITY ACCESS MANAGEMENT component to create a unique ID for each user from the student type, DATA STORE OFF CHAIN component which is an off-chain  database to store users data and certificates information in the frontend of system, HASING algorithm to create a unique hash for each uploaded certificate, TRANSACTION MANAGER to initiate and manage the transactions that send the certificate hash and the issuer meta data to the smart contract on the blockchain, WALLET to store and manage account keys, broadcast transactions, send and receive Ethereum tokens, and connect to decentralized applications, API to connect the frontend of the system to the smart contract on the blockchain to store the metadata of universities and certificates' hashes. The system's backend relies on the blockchain, a decentralized network of computer nodes that confirm and validate data added to the chain. This process results in digital data blocks being hashed and added to the chain via a cryptographic link. Blockchain-based records are reliably easy and quick to transfer, with students only needing to share a digital address to link future employers to their authenticated credentials. To integrate the blockchain with the frontend of the system, a smart contract has been written using Solidity and deployed on the Ethereum Virtual Machine (EVM) on the blockchain. The smart contract integrates on the front end through the Application Programming Interface (API).
Four actors interact with the system, the first actor is the system administrator, responsible for executing the smart contract on the blockchain and the registration of universities on the system. The second actor is the university or learning institution, responsible for the authentication of student records. The third actor is the student, who utilizes the system to create a record of their achievements. The fourth actor is an employer, who utilizes the system to validate the candidate's certification and assess candidates using their records of achievements. To illustrates the interaction with the system and defines the requirement to describe a particular use of the system, users will interact with the system in different manners.

\subsection{Implementation}
\label{Implementation}
The frontend of the system is software implemented using HTML5, CSS3, Javascript, AJAX, and Web3js. We used HTML5 to design the frontend application, By using HTML5, we made usable forms. In addition, HTML5 supports cross-platform, is designed to display the application pages on PC, Tablet, and smartphones and keeps CSS better organized. Javascript is used to allow users to interact with the system frontend and to implement the frontend components.  AJAX is used in this platform to allow a web page only to reload those portions which have changed, rather than reloading the whole page. This decentralize application is connected to a smart contract using web3js. Web3js is a collection of libraries that allows users to interact with the local or remote Ethereum node using an HTTP connection. The database has been designed to contain two categories of data: public authentication data and private certificate data. The public authentication data is available and released to the blockchain, the student data are stored in MySQL, securely protected and isolated in the intranet. The smart contract in this system is written by Solidity, a high-level and contract-oriented language used to write smart contracts. It is used for designing and implementing smart contracts. It's designed to run on the Ethereum Virtual Machine (EVM), which is hosted on Ethereum Nodes connected to the blockchain.

\subsubsection{Blockchain}
\label{Blockchain}
\begin{figure}
    \centering
    \includegraphics[width=0.48\textwidth]{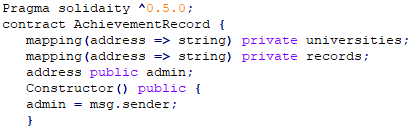}
    \caption{Achievement Record Smart Contract.}
    \label{fig:Achievement Record Smart Contract.}
\end{figure}

A blockchain-based application has been chosen for this system due to its performance and ability to verify education certifications proficiently. There are numerous reasons why blockchain is the most appropriate decision, including because it helps to remove the need for the manual verification of transactions since all necessary information is automatically verified by a decentralised network of computers. This information is also permanently stored in the blockchain, reducing, if not completely removing, the risk of deletion, meaning that additional security services are not needed. Importantly, the falsification or modification of transactions on the blockchain cannot take place. The specific hash system is used to verify certificates, and no user is capable of modifying this information or uploading a false hash into the network. 
Users, also known as nodes, can create transactions and then propagate them over the blockchain network. Miners are in charge of maintaining the blockchain network's ledger by regularly adding new blocks of transactions. Miners pick and execute a number of pending transactions from their pools to build and attach a new block to the ledger and then include them in the block by engaging in a consensus algorithm such as PoW. Subsequently, the created block will be sent to the rest of the network's nodes. When a new block is created, each node must validate it before adding it to its local copy of the blockchain. The block will be verified and deemed part of the global blockchain ledger if the majority of nodes in the network accept it, append it to their local blockchain copies, and build upon it. 
The blockchain used for this specific system is Ethereum, an open source blockchain with smart contract capabilities. It also supplies a decentralised virtual machine that can complete the necessary scripts through the use of a system of public nodes. This system is considered to be Turning complete, meaning it can recognise other data sets and is also used as the internal transaction pricing mechanism. Decentralised applications are connected to the smart contract using web3.js, which is an assortment of archives that permit the system to interact with remote or local Ethereum nodes.
The smart contract is of primary importance within the system, as it connects the blockchain with the frontend  \cite{molina2018implementation} \cite{crowcroft2018and}.
Regarding this specific platform, the use of smart contracts eliminates the need for human management as Application Programming Interface (API) connectivity instructs the smart contract; as in figure \ref{fig:Achievement Record Smart Contract.}; to execute actions on the frontend automatically. This reduces the risk of documentation fraud for universities and employers. 
The primary programming language used when writing smart contracts that run on Ethereum Blockchains is Solidity, a contract-oriented language that is responsible for the secure storage of programming logic during a transaction. Solidity is also a high-level language and is used in the design, writing, an implementation of smart contracts to run on the Ethereum Virtual Machine (EVM), which is held on Ethereum Nodes that are directly linked to the blockchain, that in turn is connected to the frontend. 
A smart contract allows this platform to register universities and store the relevant document hash values on the blockchain, with Solidity and Truffle Framework working to publish it on the Ethereum Blockchain. It employs automatically when a command is given on the frontend via API connectivity. First, universities are registered on the blockchain using an API with the formulated function add\_uni. Another formulated function, store\_hash works to store the relevant document in the smart contract and was generated by the SHA-256 encryption algorithm that subsists in the system’s frontend. The function get\_hash is used to verify particular documents on the hash through an API.

\subsubsection{Transaction and Gas}
\label{Transaction and Gas}
Register University, Store Hash, and Verify Hash are the relevant discussed transactions utilised in this particular system, and work to initiate transactions by applying the data in the transaction, the SHA-256 Hash value \cite{gueron2011sha} \cite{wood2014ethereum}, and the Ethereum Address. Once a transaction has been logged in the blockchain, the details of the transaction, including the asset, price, and ownership, are immediately confirmed within a matter of seconds throughout all nodes, with a verified alteration on one ledger being instantaneously recorded on every other ledger. A specific node on the Ethereum blockchain is used to create a wallet address, which means that, with the aid of an API, it is easy to check the balance of a fee that an admin must pay, and the mining fee is deduced automatically by the Ethereum wallet address \cite{werner2020step}.

\subsection{Users Interactions with the System}
\label{Users Interactions with the System}
\begin{figure} [!htbp]
    \centering
    \includegraphics[width=0.5\textwidth]{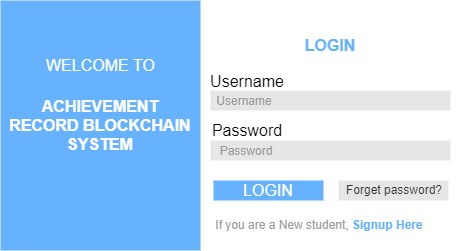}
    \caption{System Home Page}
    \label{fig:System Home Page}
\end{figure}

Users will interact with the system in different manners. Users login with the frontend function and then access and distribute their documents if they so choose; figure \ref{fig:System Home Page} shows the system home page where users can access the system. Users’ documents are uploaded on the blockchain via the smart contract, with documents also being uploaded by the university for verifiable purposes. As students can access their uploaded documents, they can send them to potential employers, and employers can also verify the document using the document hash to search the system.

\subsubsection{Admin Interaction with the system}
\label{Admin Interaction with the system}
\begin{figure}[!htbp]
    \centering
    \includegraphics[width=0.48\textwidth]{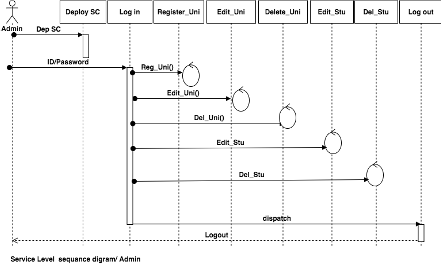}
    \caption{Admin operations sequence diagram}
    \label{fig:Admin operations sequence diagram}
\end{figure}

\begin{figure}[!htbp]
    \centering
    \includegraphics[width=0.48\textwidth]{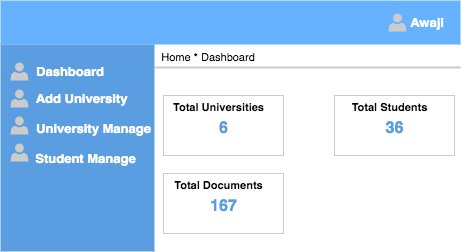}
    \caption{Admin dashboard}
    \label{fig:Admin dashboard}
\end{figure}

To begin, admins must log in to the system using the correct login credentials previously supplied to them before being forwarded to the admin dashboard, wherein the menus are laid out. Admins from their home page; as shown in figure \ref{fig:Admin dashboard}; can choose from ‘Add University’, ‘University Manage’, or ‘Student Manage’. The ‘Add University’ tab allows admins to add a university into the university database by completing the form. Later, if necessary, the ‘University Manage’ tab allows admins to edit and delete universities. The admin operations sequence presented in figure \ref{fig:Admin operations sequence diagram}.
\subsubsection{University Interaction with the system}
\label{University Interaction with the system}
\begin{figure} [!htbp]
    \centering
    \includegraphics[width=0.48\textwidth]{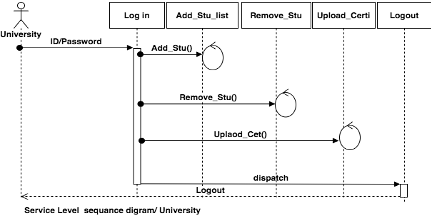}
    \caption{University operations sequence diagram}
    \label{fig:University operations sequence diagram}
\end{figure}

\begin{figure} [!htbp]
    \centering
    \includegraphics[width=0.48\textwidth]{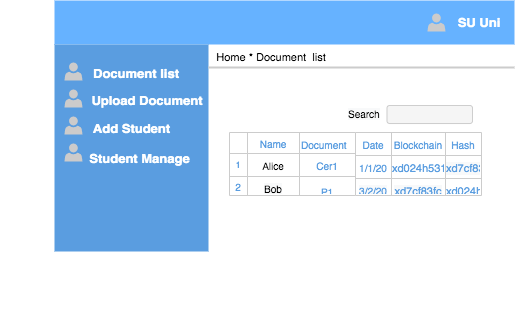}
    \caption{University Dashboard}
    \label{fig:University Dashboard}
\end{figure}

The university user must log into the system, again with correctly supplied credentials, and then will be taken to the university dashboard; as showed in figure \ref{fig:University Dashboard}. The user will be presented with specific menus, including ‘Document List’, ‘Upload Document’, ‘Add Students’, and ‘Manage Students’. By clicking on ‘Document List’, the university user will be able to view a list of the students’ documents and clicking on any specific document will produce information about the relevant student. The ‘Upload Document’ tab allows the user to upload a student’ document by entering their details. If the document type is already available in the system, they can upload the document straight away. If not, they must first add the specific document type. The ‘Add Student’ tab enables the user to add a new student and all of their relevant details into the database, and the ‘Manage Students’ tab allows the university user to see a complete list of students, editing where necessary. The university operations sequence presented in figure \ref{fig:University operations sequence diagram}.

\subsubsection{Student Interaction with the system}
\label{Student Interaction with the system}

\begin{figure}[!htbp]
    \centering
    \includegraphics[width=0.48\textwidth]{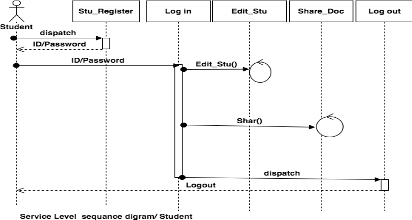}
    \caption{Student operations sequence diagram.}
    \label{fig:Student operations sequence diagram}
\end{figure}

\begin{figure}[!htbp]
    \centering
    \includegraphics[width=0.48\textwidth]{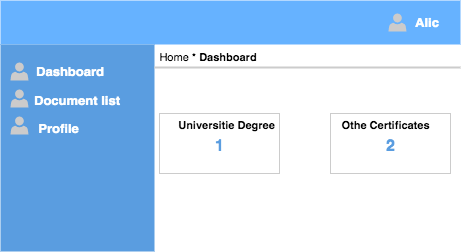}
    \caption{Student Dashboard.}
    \label{fig:Student Dashboard}
\end{figure}

The student homepage in figure \ref{fig:Student Dashboard} allows university students to register themselves in the system. They are provided with a unique user ID and password, and, once entered correctly, will be redirected to the dashboard. If they enter their details incorrectly, an error message will be displayed, and they will stay on the login page until they enter the correct details. Once successfully logged in, students can see their information on the dashboard and are also presented with a list of certificates that have been uploaded by universities. Using the email system, students can share their documents with different employers, who will receive a link directing them to a document verification page. The student operations sequence presented in figure \ref{fig:Student operations sequence diagram}.

\section{EVALUATION}
\label{EVALUATION}
To appraise our solution, we designed an experiment to collect data from the end-users of the proposed system. We mainly aim in this research to provide the stakeholders with a system that is use-friendly and trusted. Therefore, we first evaluated the system usability utilising the System Usability Scale (SUS) test~\cite{brooke1996sus}. Also, we analysed the system's feasibility in terms of cost and transaction confirmation time. We received responses from 6 universities and 30 students from those universities who agreed to participate in our proposed system's evaluation process.
\subsection{System Usability Scale (SUS)}
\label{System Usability Scale (SUS)}
To appraise our solution’s usability, we will undertake a System Usability Scale (SUS) test \cite{brooke1996sus}. The evaluation process can capture quantitative data. Usability refers to the quality of a user’s experience when interacting with the system. Usability is about effectiveness, efficiency and the overall Satisfaction of the user. The System Usability Scale (SUS) is a reliable tool for measuring usability. It consists of a 10-items questionnaire with five response options for respondents; from Strongly agree to Strongly disagree. Initially created by John Brooke in 1986, it allows the evaluation of a wide variety of products and services, including hardware, software, mobile devices, websites and applications. SUS has become an industry standard, with references in over 1300 articles and publications. 

\subsubsection{Results}
\label{Results}
The Likert scale with five steps was used to gather the responses of participants for the usability of the system. A proper process including few steps was followed to determine the response option chosen most frequently by the participants from strongly disagree to strongly agree to conclude. When asked if they think they would like to use the system more often, 36.67\% and 40\% agreed to the statement indicating that they are interested in the use. However, when asked if the system was complex, 43.33\% of participants strongly disagreed and 30\% disagreed indicating that the system was easy to use for them. It has also been found that most of them already thought that the system would be easy to use. They were then asked if they would need the support of any technical person, it was found that they do not think that they would need any kind of support while using this system, as 30\% and 43.33\% disagreed and strongly disagreed with the statement. The result also indicated that participants think that the system has integrated its various functions. They disagreed with the question that there was inconsistency in the system and agreed with the statement that people can learn using this system quickly. 63.33\% and 23.33\% strongly disagreed and disagreed with the statement that the system is burdensome to use, and they were also confident about using the system. Participants also disagreed with the statement that they had to learn a lot of things before using this system.
The overall analysis indicated that participants found the system very facilitating and easy to use, they believe that there is no need for any additional learning and neither does the system want any expertise to be used. Participants found the system user-friendly and highly useful which indicates that participants were happy about the use of this system. The overall results show that the participants are satisfied with the usability of the system and find it user-friendly.

\begin{figure}
    \centering
    \includegraphics[width=0.48\textwidth]{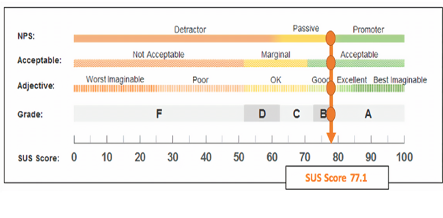}
    \caption{The SUS Scale Obtained.}
    \label{fig:The SUS Scale Obtained}
\end{figure}

\subsubsection{Discussion}
\label{Discussion}
The effectiveness of a system is usually assessed by the feedback of its users as to whether the system 
facilitated them, how easy it was to use the system and how much time and effort did the system take for understanding. It has been found that the system was highly useful for them in terms of ease of use due to which they think they will like to use the system frequently. 36.67\% and 40\% stated that they would like to use the system more frequently. Moreover, 43.33\% of participants strongly disagreed and 30\% disagreed when asked if the system was complex. It has also been found that they do not feel dependent upon anyone for using the system, for example, no expertise is required to use the system which shows that the system is very user-friendly and this feature is considered one of the most important considerations to declare a system effective and usable, 30\% and 43.33\% disagreed and strongly disagreed to the statement that the system needs some expertise to use. The responses of participants have stated that they are satisfied with the usability and efficiency of the system while it has also been found that the system can be learned very quickly without any training or expert support. Neither is the system found to be cumbersome and the participants felt very confident while using the system. Overall, participants stated that there was no hardship they faced while using the system as it does not require any expertise or training to use. Hence, anybody can easily learn how to use the system. Figure~\ref{fig:The SUS Scale Obtained} that shows a scale 77.1 which falls under good and excellent to rate the system. It is grade B which is a positive indication and rating based on the SUS scale Interpretation.

\subsection{Transactions Confirmation Time and Transactions Cost}
\label{Transactions Confirmation Time and Transactions Cost}
This analysis discusses two variables:

\textbf{Delay time}: Delay time represents the transaction confirmation time. It refers to the time a transaction takes from its broadcast to the blockchain and addition to the distributed ledger.

\textbf{Transaction cost}: The cost represents the mining fee (Gas). Gas refers to the unit that measures the computational effort required to execute specific operations on the Ethereum network. Gas fees are paid in Ethereum’s native currency, Ether (ETH).

\subsubsection{Results}
\label{Results2}
The collected data revealed 219 transactions taking approximately 17 minutes and 32 seconds to confirm. The average time to confirm the transaction was 0.24 minutes. The data shown in table~\ref{table:Transaction Times and Costs.} represents the total transaction fee measured as 00.767 Ether, however, the average transaction fee stood at 0.01097 Ether. On evaluating the transaction fee in USD, the attained rate stood at \$657.409. Similar to this, the average transaction fee in USD stood at \$6.1574. Thus, the table \ref{table:Transaction Times and Costs.} below shows the overall summary of the transactions discussed in the study.

\begin{figure}[!htbp]
    \centering
    \includegraphics[width=0.46\textwidth]{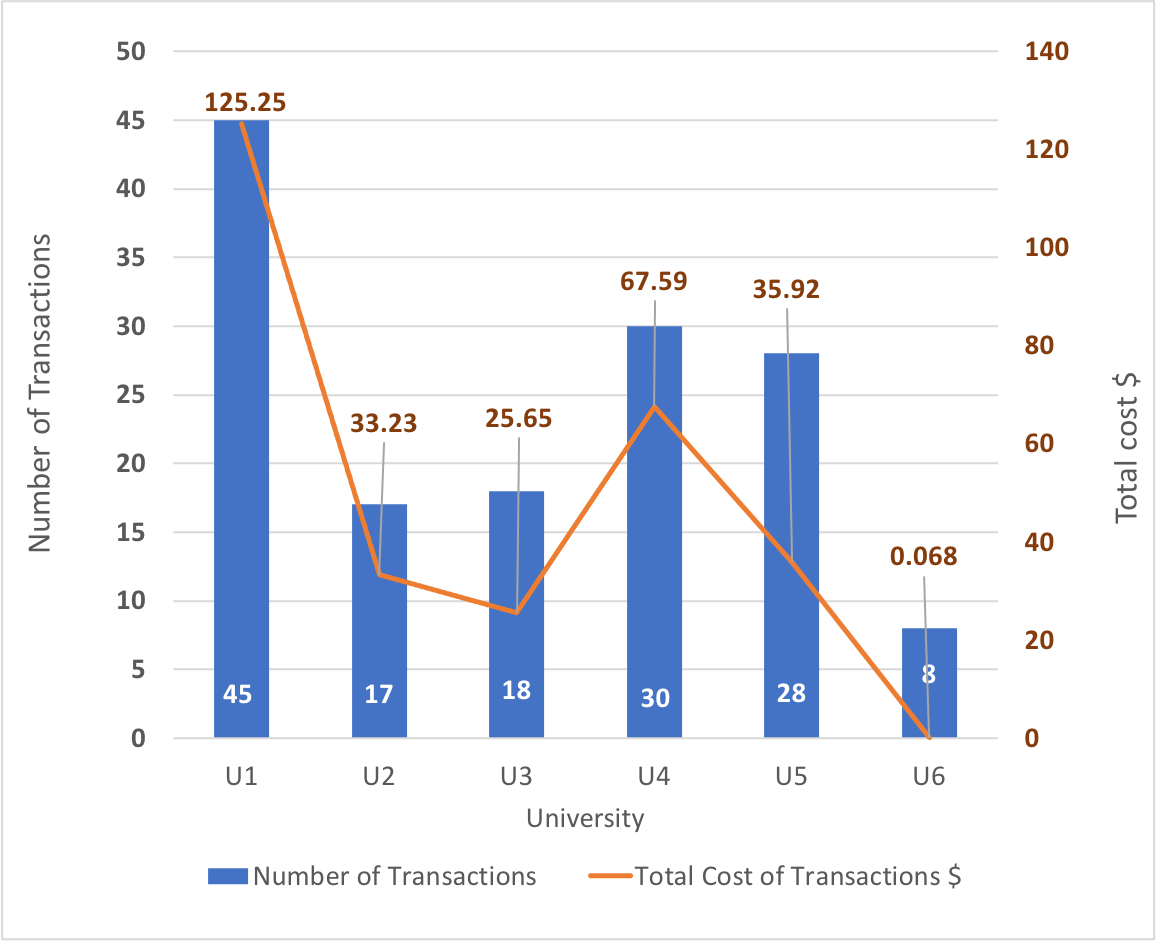}
    \caption{Number of Transactions and Total Cost (US Dollar \$)}
    \label{fig:Number of Transactions and Total Cost (US Dollar)}
\end{figure}

\begin{figure}[!htbp]
    \centering
    \includegraphics[width=0.46\textwidth]{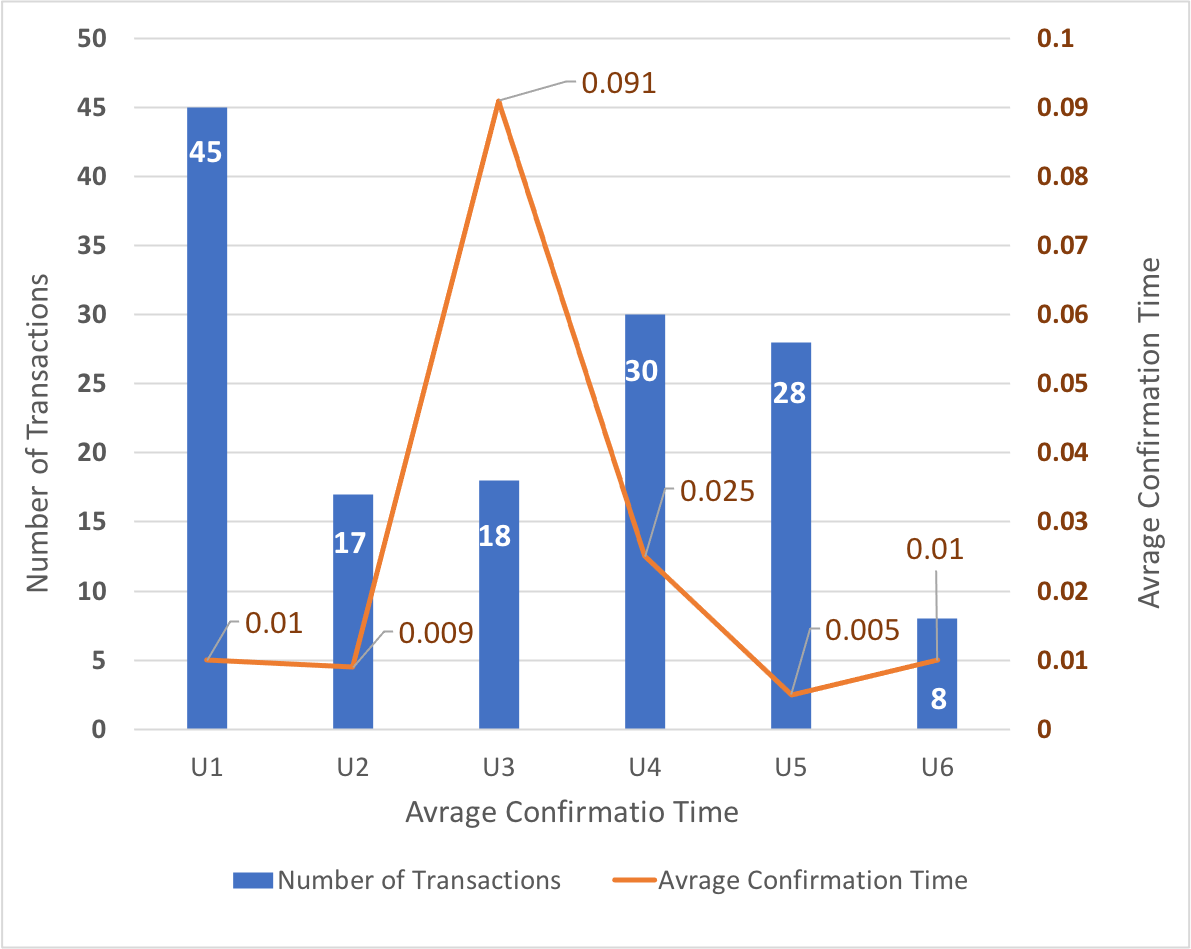}
    \caption{Number of Transactions and Average Confirmation Time.}
    \label{fig:Number of Transactions and Average Confirmation Time.}
\end{figure}

\begin{figure}[!htbp]
    \centering
    \includegraphics[width=0.46\textwidth]{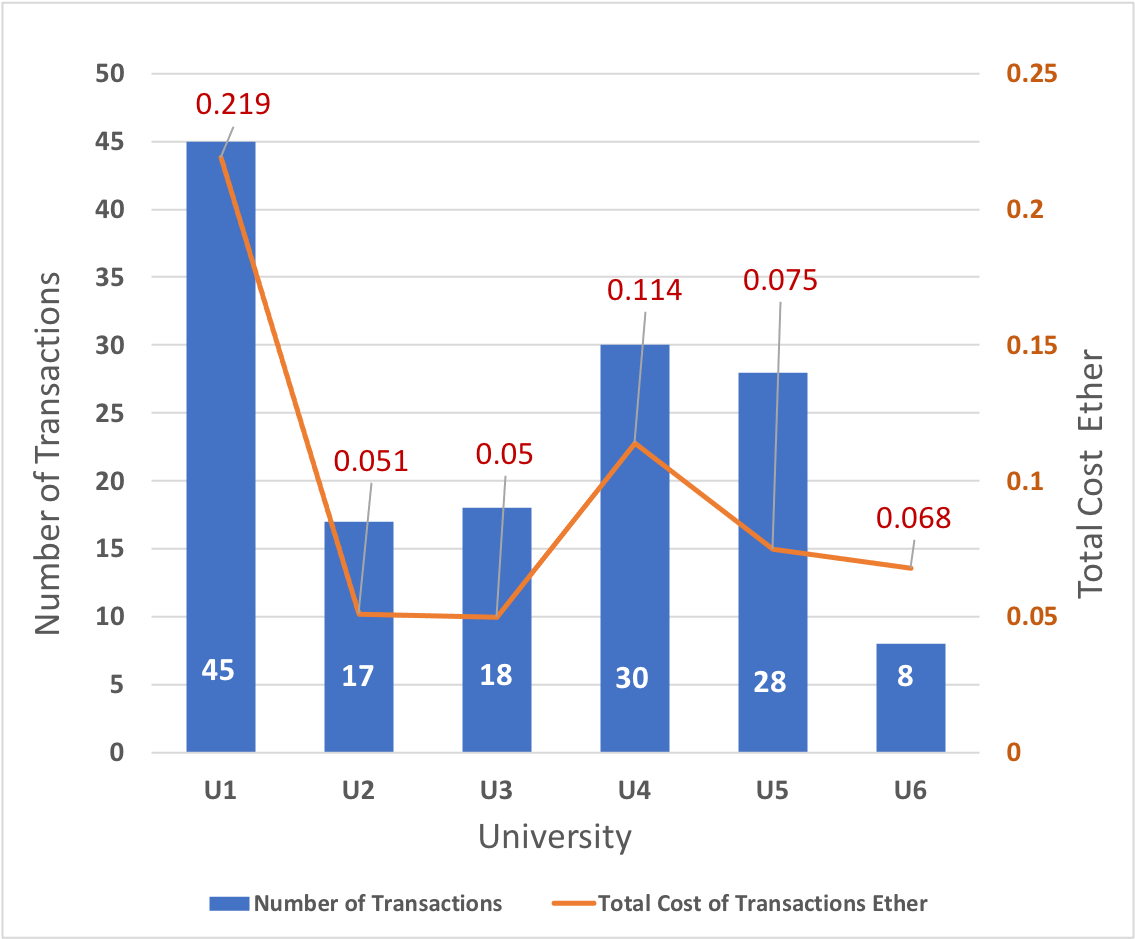}
    \caption{Number of Transactions and Total Cost (Ether).}
    \label{fig:Number of Transactions and Total Cost (Ether)}
\end{figure}

\begin{table}[!htbp] 
\centering
\caption{Transaction Times and Costs.}
\label{table:Transaction Times and Costs.}

\begin{tabular}{|l|l|} 
\hline
Total Number        & 219              \\
     of Transactions &              \\ \hline
Total Confirmation    & 17:32            \\ 
Time (MM: SS)  &  \\\hline
Average Confirmation  & 0.24             \\ 
Time (MM: SS) & \\ \hline
Total Transactions     & 0.76703077  \\ 
Fee (Ether) & Ether\\ \hline
Average  Transactions   & 0.01091671  \\ 
Fee (Ether) & Ether\\ \hline
Total Transactions       & \$657.409294     \\ 
Fee (USD) & \\ \hline
Average  Transactions     & \$6.15745258     \\ 
Fee (USD) & \\ \hline
\end{tabular}
\end{table}
Each transaction has also been calculated using different universities to obtain more complete results. In total, six universities are targeted. According to the data obtained, the total number of transactions for University 1 was 45 transactions, whereas the confirmation time was 10 minutes 5 seconds. Accordingly, the transaction fees were 0.21992263 Ether. Conversely, the average transaction fee was 0.00549 Ether.
Data from University 2 revealed that 17 transactions occurred. Accordingly, the result exhibited that the total confirmation time was 3 minutes 44 seconds, although the average confirmation time was 14 seconds. Moreover, the transaction fee was in Ether, with the average transaction fee being 0.00319 Ether. 
Figure~\ref{fig:Number of Transactions and Total Cost (US Dollar)} illustrates the number of transactions while comparing them with an average confirmation time. University 1 had 45 transactions. University 2 had 17 transactions, whereas University. 
3 had 18. University 4 handled 30 transactions, whereas universities five and six had 28 and 8 transactions, respectively. Universities 1 and 6 took the same time. The number of transactions for University 1 was 45, whereas University 6 had just 8. The assessment depends on not only the time spent but also the number of transactions. The difference is significant. University 2 took 0.009 seconds and performed 17 transactions, University 3 carried out 18 transactions in 0.091 seconds, University 4 undertook 30 transactions in 0.025 seconds, and University 5 completed 28 transactions in 0.005 seconds. The highest number of transactions was by University 1, with 45 in the least time taken by any universities.
The study established that all universities deal with different transaction costs (Ether). University 1 had the highest such costs with 0.219 Ether, whereas University 2 had the lowest with a rate of 0.051 Ether. The total transaction cost obtained from University 4 was measured as 0.114 Ether. Meanwhile, University 3 had a transaction cost of 0.05 Ether. The data analysis compared the number of transactions and total cost (Ether). The findings confirm that University 1 had the highest number of transactions while dealing with the costs of 0.219 Ether. The number of transactions resulting from University 2 stood at 17, however, its total cost reached 0.051 Ether. Similarly, the total transaction cost for University 4 was 0.114 Ether, whereas the number of transactions totals 30. The cost of Ether varies according to the number of transactions, with University 1 having the most Ether and University 3 the least. Although universities 3 and 6 had approximately the same Ether costs, the number of transactions differed. For instance, University 3 made 18 transactions, whereas University 6 performed 8. University 2 made 17 transactions in 0.051 seconds, regarding Ether cost, whilst University 6 conducted the lowest number of transactions.
Accordingly, the lowest ratio was associated with University 6, which had 8 transactions. However, the total cost of transactions in US dollars was measured at \$0.068. All universities had different amounts of transactions and average costs, as reflected in the graph below. The number of transactions at University 1 amounted to 45, and the total cost generated by the university was \$125.25. The highest cost of the transaction generated by the universities. The cost is high because the university made the highest number of transactions in the least time. The lowest transactions, 8, were made by University 6. The cost was low with 0.068, because the institution made the fewest transactions simultaneously as University 1. Here, the study determines the time and amount of transactions in a certain time count when generating costs, as shown in figures~\ref{fig:Number of Transactions and Average Confirmation Time.} and~\ref{fig:Number of Transactions and Total Cost (Ether)}. 

\subsubsection{Discussion}
\label{Discussion2}
These findings reveal different average transaction confirmation times regarding blockchain for each university. Similarly, the study notes a difference in the cost of transactions for each transaction on the blockchain. Therefore, cost and time to confirm the transaction represent the variables considered for the discussion.

\textbf{Delay Time}: Delay time represents the transaction confirmation time. Specifically, the time a transaction takes from its broadcast to the blockchain is added to the distributed ledger. From the quantitative data analysis in the previous section, the findings indicate that for 219 transactions, there exists a fluctuation in the average transaction confirmation time between the universities. Ethereum network congestion might represent the cause of delay in the transaction. 
In terms of transaction confirmation times, the analysis reveals that the average confirmation time proved different among the universities. This situation indicates the transactions’ time variations can evaluate the system’s efficiency when added to the blockchain. By considering the output provided in the previous section, the study ascertained that the number of transactions handled by University 3 was only 18. However, it took 0.09 minutes on average to complete the transaction. Meanwhile, University 3 had the most transactions, but only 0.01 minutes on average for transaction confirmation. Hence, the average confirmation time does not depend upon the number of transactions handled by the system but the efficiency of the blockchain system and congestion on the Ethereum blockchain. When a blockchain network experiences peak traffic, it delays transactions. Other factors may also delay transaction confirmation, such as the gas limit. There exists a proportionality between the gas limit determined by the sender and the blockchain mining process. Gas prices are denoted in Gwei, which itself is a denomination of Ether. Transactions with higher gas limits attract miners, therefore, operations with lower gas limit values will continue waiting.

\textbf{Transaction Cost}: The cost is the mining fee, ‘Gas’, which refers to the unit that measures the computational effort required to execute specific operations on the Ethereum network. The purpose of gas is to control the resources that a transaction can use since it will be processed on computers worldwide. Gas is separate from ETH to protect the system from volatility in the value of ETH and manage the ratios between the costs of various resources that gas pays for, such as computation, memory and storage. Gas also rewards the miners for the work they do. The gas price component of a transaction allows users to set the price they want to pay in exchange for gas, where the price is measured in Gwei per gas unit. Wallets can change the gas price to achieve faster transaction confirmations—the greater the gas price, the quicker the transaction confirmation. Accordingly, the gas limit for the transactions sent from the proposed system stood at 40,000 Gwei (0.004 ETH). Concerning the volume of data sent in each transaction, this value proves attractive to miners. Consequently, all user transactions during the system evaluation required confirmation in a good average time. Lower priority transactions can use a lower gas price which means a slower confirmation. The market decides the relationship between the price of ETH and the cost of computing operations concerning gas. The gas cost acts as a measure of computation and storage used in the EVM, where the gas has a price measured in Ether. When sending a transaction, people can specify the gas price they want to pay in ETH for each gas unit. This equation calculates the transaction fee:

\begin{multline}
\mbox{Transaction fee} = \mbox{total gas used} \\ \times \mbox{gas price paid (in Ether)} 
\end{multline}

This study can explain why each transaction fee differs. Both factors in the equation play a role in determining the cost of transitions.\newline

\textbf{Total gas used:} the gas limit has been specified in the system for each transaction at the value of 40,000 Gwei (0.004 ETH) to speed up the mining process. It remains unnecessary to use the specified gas limit since transactions must pay for the computational, bandwidth and storage space they consume in proportion to these gas costs. Although a transaction includes a limit, any gas not used in a transaction reverts to the user. In this sense, the value of ‘total gas used’ changes with each transaction and, therefore, the fee changes accordingly.
    
\textbf{Gas price paid (in Ether):} gas fees are paid in Ethereum’s native currency (ETH). Gas prices are denoted in Gwei. A Gwei or Gigawei is defined as 1,000,000,000 Wei, the smallest base unit of Ether. One Gwei equals 0.000000001 or 10-9 ETH. Conversely, 1 ETH represents 1 billion Gwei. Consequently, each cost in Ether is different due to the constant change in the value of Ether in the stock market. For example, when this study started the system evaluation process in May 2020, the price of Ether began at \$213.61, meaning that the transaction fee was 0.004$\times$213.61 = \$0.854. Meanwhile, the transaction fee in February 2021 was 0.004$\times$2036.55 = \$8.1462. The average fee for transactions. This study also noted the significant difference between the cost in less than a year, with the price of Ether more than tripling. Thus, the transaction cost doubled in proportion to the rise of Ether in the price of cryptocurrencies in the stock market.
Considering this aspect and the information provided in the previous section, it remains difficult to ascertain accurate confirmation times when sending a transaction from one node to another and adding it to the distributed ledger on the blockchain. Furthermore, the transaction fee remains unfixed and cannot be determined until confirmation of the transaction. Such a situation depends on different factors. Based on previous findings and the points discussed in this section, the irregular transaction time depends primarily on the efficiency of the network but remains within the acceptable range. However, this area represents one of the limitations requiring evaluation in future research. The unfixed transaction cost changes depending on crypto prices on the stock market. This situation may affect the sustainable use of the system. When the cost proves too high, the system will be less attractive to users. Thus, transaction fees represent another significant restriction of the system. Despite the effectiveness in other regards and its achievement of the research objectives, the cost and the scalability of blockchain and energy consumption are negative aspects that open the door for future research. In our system, we have to execute various functions on the blockchain, making the smart contract an essential component in the system structure. Therefore, the blockchain platforms that do not support smart contracts are ineffective in tackling this issue.
Table \ref{table:Different Blockchain Platforms and Their Characteristics.} illustrates different blockchain platforms, each has different characteristics and design decisions. The blockchain platforms listed in table~\ref{table:Different Blockchain Platforms and Their Characteristics.}, Ethereum and Hyperledger, are the only platforms designed to support rich and complex smart contracts. Ethereum is a permissionless blockchain platform designed to support creating and deploying complex smart contracts on blockchains. While Hyperledger is an open-source collaborative project aiming to advance permissioned blockchains, it aims to provide an infrastructure of different modules, such as smart contract engines, and tools for developing blockchain platforms.

\begin{table}[!htbp] \small
    \centering
    \caption{Different Blockchain Platforms and Their Characteristics.}
    \label{table:Different Blockchain Platforms and Their Characteristics.}
    \begin{tabular}{|c|c|c|}
    \hline
        Blockchain & Network &Smart Contract  \\
         Platform&  Permission&  Support\\ \hline
         Bitcoin     & Permissionless              & No  \\ \hline
        Ethereum    & Permissionless              & Yes \\ \hline
        Zcash       & Permissionless              & No  \\ \hline
        Litecoin    & Permissionless              & No  \\ \hline
        Dash        & Permissionless              & No  \\ \hline
        Peercoin    & Permissionless              & No  \\ \hline
        Ripple      & Permissionless& No  \\ 
        & (controlled) &   \\ \hline
        Monero      & Permissionless              & No  \\ \hline
        MultiChain  & Permissionled               & No  \\ \hline
        Hyperledger & Permissionled               & Yes \\ \hline 
    \end{tabular}
\end{table}

In theory, developers can use Ethereum or Hyperledger to build the distributed application based on the system requirements and objectives. However, integrating the Hyperledger Fabric platform instead of the Ethereum platform into such a system is an interesting project for future researchers to tackle the transaction cost issue. From a different perspective, a private Ethereum blockchain may be an appropriate solution to the transaction cost. It comprises zero transaction fees and higher scalability, and there are no restrictions. However, switching from public to private blockchain requires a range of additions and modifications in the design of distributed applications. When compared against public blockchain, private blockchain nodes require permission to join a controlled blockchain and read the chain’s state. Only users with permission can subscribe to the network and write or send transactions to the blockchain. Therefore, converting to a private blockchain is another possible solution for future researchers to tackle the issue. 

\begin{table}[!htbp] \small
    \centering
    \caption{Financial Comparison of CVSS and Our System.}
    \label{table:Financial Comparison of CVSS and Our System.}
    \begin{tabular}{|c|c|c|}
    \hline
                          & CVSS & Our System \\\hline
    Contract Creation Cost& \$19 & \$10.76 \\ \hline
    Number of Transactions& 60 & 219 \\ \hline
    Transaction Cost & \$0.15 & \$6.16 \\ \hline
    Average Transaction & 00:60 & 00:24\\ 
    Confirmation Time (mm:ss)& & \\ \hline
     Total Transactions & 05:00 & 17:32\\ 
    Confirmations Time (mm:ss)& & \\ \hline
    \end{tabular}
\end{table}
CVSS system is the only system that conducted financial analytic transactions, provided an explanation of the transaction cost, while also providing the time of confirmation of transactions on the Ethereum blockchain. Therefore, we compared the financial analysis of our system with the CVSS system, as shown in table~\ref{table:Financial Comparison of CVSS and Our System.}. According to the principle of transaction cost, the cost of deploying the smart contract on the Ethereum blockchain for the CVSS system is \$19, whereas it is \$10.76 for our proposed system because of the optimisation we carried out in the smart contract. Accordingly, it only contains those functions that are necessary to be on the blockchain. The number of transactions conducted in our proposed system is three times greater than in CVSS. This provides a clearer perception of the transaction cost when using the system. Furthermore, we observe that the cost of a single transaction in the CVSS system is \$0.15, while no clarification is provided as to whether this number is fixed per transaction, or that it is the average cost for 60 transactions sent through the CVSS system. 
However, we calculated the mean cost for 219 transactions in our system, due to these transactions being sent at various times over a period greater than six months, during the system evaluation carried out by end-users. The mean transaction cost for those sent from our proposed system is \$6.16. We can explain the cost rise as being a consequence of the rapid rise in ether’s price during the system evaluation period carried out by end-users, as explained previously in the evaluation section. 
Regarding the principle of transaction confirmation time, the mean transaction confirmation time in the CVSS system was 60 seconds, whereas the average for our proposed system was 24 seconds, which is deemed acceptable in contrast with CVSS. The reason for this may be the gas limit or the propagation delay due to Ethereum network congestion, as clarified in the evaluation section. Additionally, the total transaction confirmation time in the CVSS system for 60 transactions was five minutes. Comparatively, the total transaction confirmation time in our proposed 219 transaction system was approximately 17 minutes, which is an acceptable number given the total number of transactions. 

\section{CONCLUSIONS}
\label{CONCLUSIONS}
This research has aimed to demonstrate that the use of Blockchain technology for the certification and verification of achievements in higher education has great potential in the global market. It is a method that would be sustainable and advantageous for multiple parties. Credential fraud is widespread and pervasive, having a negative impact on educational institutions, students, and the wider society. Current solutions, such as legacy credential verification systems, are clumsy and are not time nor cost-efficient. In addition to this, they lack efficacy in their response to corrupt practices, such as fraud on the part of educational institutions and accreditation bodies. The record of achievements that is proposed with the use of Blockchain technology is comprehensive in tackling widespread fraud. This system is significantly improved in comparison to legacy systems, being both more user-friendly and more efficient. The Blockchain technology method is a solution that effectively integrates into the existent credential verification ecosystem. This work aspires to positively contribute to ongoing efforts towards the prevention of credential fraud. Nevertheless, the cost of transactions and energy consumption because of the PoW algorithm are essential aspects to consider during the conception of a blockchain-based solution.

\section*{Acknowledgements}
We are grateful to all of those we have had the pleasure to work with during this project and all participants who contributed to validating the design. In addition, we thank all the participants who answered the questionnaires and others who participated in the evaluation and reviewing process.

\bibliographystyle{apalike}
{\small
\bibliography{References}}
\end{document}